\documentclass[aps, prd, onecolumn, groupedaddress, nofootinbib, preprintnumbers, 11pt]{revtex4-2}

\usepackage{radiative_semileptonic_B_decays}

\begin{document}

\title{Radiative Semileptonic \texorpdfstring{$\bar{B}$}{TEXT} Decays}

\author{Michele Papucci}
\email{mpapucci@caltech.edu}
\author{Tanner Trickle}
\email{ttrickle@caltech.edu}
\author{Mark B. Wise}
\email{wise@caltech.edu}
\affiliation{Walter Burke Institute for Theoretical Physics, California Institute of Technology, Pasadena, CA 91125, USA}

\preprint{CALT-TH-2021-036}

\begin{abstract}

\center {{\bf \large Abstract}}

We consider the form factors for the radiative semileptonic decays $\bar{B}(v) \rightarrow D^{(*)}(v') \ell {\bar \nu}_\ell \gamma$ in the kinematic region where the photon momentum, $k$, is small enough that heavy quark symmetry (HQS) can be applied without the radiated photon changing the heavy quark velocity (i.e., $v^{(\prime)} \cdot k < m_{(b,c)}$). We find that HQS is remarkably powerful, leaving only four new undetermined form factors at leading order in $1/m_{(b,c)}$. In addition, one of them is fixed in terms of the leading order Isgur-Wise function in the kinematic region, $v^{(\prime)}\cdot k < \Lambda_\text{QCD}$.
\end{abstract}

\maketitle
\newpage
\tableofcontents
\newpage

\section{Introduction}
\label{sec:introduction}

In the limit where the bottom and charm quark are taken as much heavier than the scale of the strong interactions, QCD contains a heavy quark spin-flavor symmetry (or simply heavy quark symmetry (HQS)) \cite{Nussinov:1986hw, Shifman:1987rj, Isgur:1989vq, Isgur:1990yhj} that acts on heavy quarks with the same four velocity. These heavy quark spin-flavor symmetries are a powerful tool for understanding some non-perturbative properties of hadrons containing a heavy quark. One of the earliest, and still most important, applications is to the form factors for the semileptonic decays of ${\bar B}$ mesons to $D$ and $D^*$ mesons, i.e., ${\bar B} \rightarrow D^{(*)} \ell {\bar \nu}_\ell$, where $\ell \in \{ e, \mu, \tau \} $. Lorentz symmetry and parity invariance of the strong interactions imply that these decays are characterized by six Lorentz invariant form factors, but HQS implies that there is just one (the Isgur-Wise function~\cite{Isgur:1990yhj}). Furthermore it is normalized to unity at the zero recoil point where the final state charmed meson is at rest in the rest frame of the decaying ${\bar B}$ meson. In addition the matrix elements of the weak vector and axial vector current are independent of the masses of the heavy quarks. 

Radiative semileptonic ${\bar B}$ decays to charmed mesons provide a unique laboratory to investigate the predictions of HQS in more complicated kinematic settings, while keeping one initial and one final heavy meson as the only hadrons in the process. The presence of the extra photon provides new kinematic invariants beyond $w \equiv v \cdot v'$, namely $v\cdot k$ and $v' \cdot k$. Here $v$ is the four velocity of the decaying ${\bar B}$ meson, $v'$ is the four velocity of the final state $D^{(*)}$ meson, and $k$ is the four momentum of the final state photon. 

Belle II should allow for the study of radiative semileptonic ${\bar B}$ decay at a level of precision approaching what the previous generation of $B$ factories, i.e., Belle~\cite{Belle:2015pkj, Belle:2015qfa, Belle:2016dyj, Belle:2016ure, Belle:2017ilt, Belle:2018ezy, Belle:2019gij, Belle:2019rba}, BaBar~\cite{BaBar:2007ddh, BaBar:2007cke, BaBar:2007nwi, BaBar:2008zui, BaBar:2009zxk, BaBar:2012obs, BaBar:2013mob}, as well as LHCb~\cite{LHCb:2015gmp, LHCb:2017rln, LHCb:2017smo}, did for the semileptonic ${\bar B}$ decays. 

Moreover, radiative semileptonic ${\bar B}$ decays are an irreducible background to the measurement of non-radiative semileptonic measurements, in the kinematic regions where the photon is not reconstructed. These kinematic regions are dominated by the soft region for Belle II and are more complicated for LHCb, due to the detector geometry and the boost distribution of the ${\bar B}$ mesons. It has been shown~\cite{Bernlochner:2010fc,Cali:2019nwp} that differences in modeling the radiation processes may induce few percent systematic uncertainties in the semileptonic measurements. Having good control of radiative semileptonic processes is therefore necessary for precision studies of the non-radiative semileptonic decays by Belle II and LHCb.

In the $k \rightarrow 0$ limit, the radiative and non-radiative processes are related by gauge invariance. Previous work on radiative semileptonic $\bar B$ decays has mostly focused in the regime where the photon is soft ($k < \Lambda_\text{QCD}$)~\cite{Becirevic:2009fy,Bernlochner:2010yd,deBoer:2018ipi}. Away from the soft limit, uncalculable ``structure dependent'' (SD) contributions arise~\cite{Gasser:2004ds}, stemming from the composite nature of the $\bar{B}$ and $D^{(*)}$ mesons. Lorentz invariance and parity invariance of the strong interactions imply that the radiative decays are characterized by 32 scalar form factors that can depend on $v\cdot k$, $v' \cdot k$ and $w$; a daunting number for gaining knowledge of these SD terms either from experimental measurements or lattice QCD calculations. However, similar to the case of non-radiative semileptonic ${\bar B}$ decays, HQS greatly reduces the number of these form factors.

The purpose of this paper is to initiate this investigation and work out the predictions of HQS for radiative semileptonic ${\bar B}$ decays. Here we will work to leading order in the heavy quark mass expansion, neglecting terms suppressed by powers of $1/m_{b,c}$. Furthermore we work in the kinematic region where radiating the photon does not change the heavy quark velocities. It is in this kinematic region that HQS is most powerful. We find that with these approximations there are only four new (i.e., not proportional to the Isgur-Wise function) Lorentz invariant form factors, and one is normalized to the Isgur-Wise function as $k \rightarrow 0$.

In Sec.~\ref{sec:matrix_element} we begin by calculating the matrix element, $\mathcal{M}$, in terms of general hadronic form factors for $\bar{B} \rightarrow D^{(*)} \ell \bar{\nu}_\ell \gamma$ using heavy quark effective theory~\cite{Politzer:1988bs, Eichten:1989zv, Eichten:1990vp, Georgi:1990um}. Then, in Sec.~\ref{sec:form_factors}, we reduce the number of new, independent form factors to just four, $\zeta_{1-4}$, using the Ward identities. In Sec.~\ref{sec:soft_limit} we explore the soft photon limit including the leading order behavior in Sec.~\ref{subsec:leading_soft} and some terms at sub-leading order in Sec.~\ref{subsec:subleading_soft}. A concluding discussion is given in Sec.~\ref{sec:conclusion}. Two appendices are included; the first, Appendix~\ref{app:expanded_form_factors} works out the general structure of the matrix element, $\mathcal{M}$, for $\bar{B} \rightarrow D^{(*)} \ell \bar{\nu}_\ell \gamma$ imposing only Lorentz invariance and the parity symmetry of the strong interactions. The second, Appendix~\ref{app:other_matrix_elements}, generalizes the results to new interactions that do not have the usual $V-A$ structure of the weak interactions; potentially useful in the future for new physics signatures.

\section{Matrix Element}
\label{sec:matrix_element}

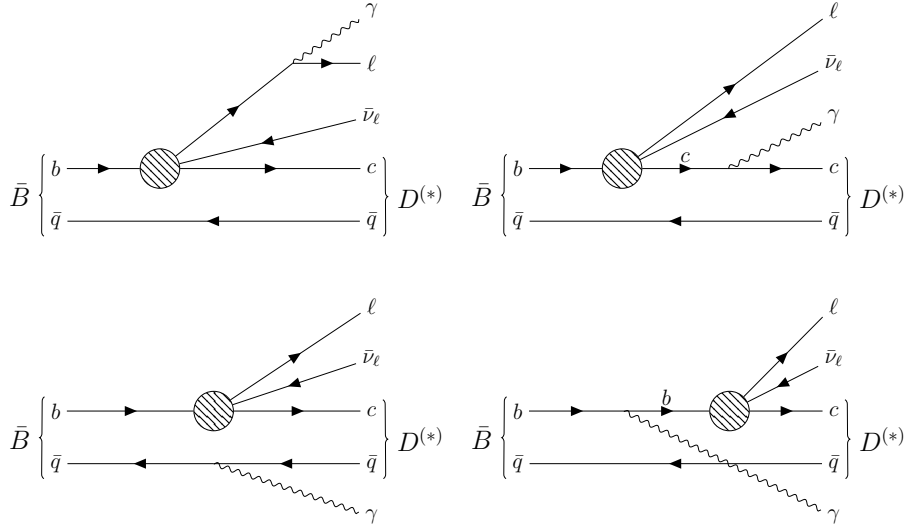
\begin{figure*}[t]
    \centering
    \scalebox{0.7}{\begin{tikzpicture}
    \begin{feynman}
        \vertex at (0, 0) (b) {\large \( b \)};
        \vertex at (6cm, 0) (c) {\large \( c \)};
        \vertex[blob] (weakInt) at ($(b)!0.33!(c)$) {};
        \vertex at ($(b)!0.5!(c) + (0, 2cm)$) (lvVertex);
        \vertex[above=1cm of c] (v) {\large \( \bar{\nu}_\ell \)};
        \vertex[above=2cm of c] (l) {\large \( \ell \)};
        \vertex at ($(l)!0.5!(lvVertex)$) (emVertex);
        \vertex[above=1cm of l] (photon) {\large \( \gamma \)};
        \vertex[below=1cm of b] (d1) {\large \( \bar{q} \)};
        \vertex[below=1cm of c] (d2) {\large \( \bar{q} \)};
        \diagram* {
            (b) -- [fermion] (weakInt) -- [fermion] (c),
            (v) -- [fermion] (weakInt),
            (weakInt) -- [fermion] (emVertex) -- [fermion] (l),
            (emVertex) -- [boson] (photon),
            (d2) -- [fermion] (d1)
        };
        \draw [decoration={brace}, decorate] (d1.south west) -- (b.north west)
            node [pos=0.5, left] {\Large \( \bar{B} \;  \)};
        \draw [decoration={brace}, decorate] (c.north east) -- (d2.south east)
            node [pos=0.5, right] {\Large \( \; D^{(*)} \)};
    \end{feynman}
\end{tikzpicture}}
    \scalebox{0.7}{\begin{tikzpicture}
    \begin{feynman}
        \vertex at (0, 0) (b) {\large \( b \)};
        \vertex at (6cm, 0) (c) {\large \( c \)};
        \vertex[blob] at ($(b)!0.33!(c)$) (weakInt) {};
        \vertex at ($(b)!0.5!(c) + (0, 2cm)$) (lvVertex);
        \vertex[above=2cm of c] (v) {\large \( \bar{\nu}_\ell \)};
        \vertex[above=3cm of c] (l) {\large \( \ell \)};
        \vertex[below=1cm of b] (d1) {\large \( \bar{q} \)};
        \vertex[below=1cm of c] (d2) {\large \( \bar{q} \)};
        \vertex at ($(weakInt)!0.5!(c)$) (emVertex);
        \vertex[above=1cm of c] (photon) {\large \( \gamma \)};
        \diagram* {
            (b) -- [fermion] (weakInt) -- [fermion, edge label=\large\( c \)] (emVertex) -- [fermion] (c),
            (v) -- [fermion] (weakInt),
            (weakInt) -- [fermion] (l),
            (emVertex) -- [boson] (photon),
            (d2)  -- [fermion] (d1)
        };
        \draw [decoration={brace}, decorate] (d1.south west) -- (b.north west)
            node [pos=0.5, left] {\Large \( \bar{B} \;  \)};
        \draw [decoration={brace}, decorate] (c.north east) -- (d2.south east)
            node [pos=0.5, right] {\Large \( \; D^{(*)} \)};
    \end{feynman}
\end{tikzpicture}}

    \vspace{2em}

    \scalebox{0.7}{\begin{tikzpicture}
    \begin{feynman}
        \vertex at (0, 0) (b) {\large \( b \)};
        \vertex at (6cm, 0) (c) {\large \( c \)};
        \vertex[blob] at ($(b)!0.5!(c)$) (weakInt) {};
        \vertex at ($(b)!0.5!(c) + (0, 2cm)$) (lvVertex);
        \vertex[above=1cm of c] (v) {\large \( \bar{\nu}_\ell \)};
        \vertex[above=2cm of c] (l) {\large \( \ell \)};
        \vertex[below=1cm of b] (d1) {\large \( \bar{q} \)};
        \vertex[below=1cm of c] (d2) {\large \( \bar{q} \)};
        \vertex at ($(d2)!0.5!(d1)$) (emVertex);
        \vertex[below=1cm of d2] (photon) {\large \( \gamma \)};
        \diagram* {
            (b) -- [fermion] (weakInt) -- [fermion] (c),
            (v) -- [fermion] (weakInt),
            (weakInt) -- [fermion] (l),
            (emVertex) -- [boson] (photon),
            (d2) -- [fermion] (emVertex) -- [fermion] (d1)
        };
        \draw [decoration={brace}, decorate] (d1.south west) -- (b.north west)
            node [pos=0.5, left] {\Large \( \bar{B} \;  \)};
        \draw [decoration={brace}, decorate] (c.north east) -- (d2.south east)
            node [pos=0.5, right] {\Large \( \; D^{(*)} \)};
    \end{feynman}
\end{tikzpicture}}
    \scalebox{0.7}{\begin{tikzpicture}
    \begin{feynman}
        \vertex at (0, 0) (b) {\large \( b \)};
        \vertex at (6cm, 0) (c) {\large \( c \)};
        \vertex[blob] at ($(b)!0.67!(c)$) (weakInt) {};
        \vertex at ($(weakInt)!0.5!(c) + (0, 1.5cm)$) (lvVertex);
        \vertex[above=1cm of c] (v) {\large \( \bar{\nu}_\ell \)};
        \vertex[above=2cm of c] (l) {\large \( \ell \)};
        \vertex[below=1cm of b] (d1) {\large \( \bar{q} \)};
        \vertex[below=1cm of c] (d2) {\large \( \bar{q} \)};
        \vertex at ($(b)!0.5!(weakInt)$) (emVertex);
        \vertex[below=1cm of d2] (photon) {\large \( \gamma \)};
        \diagram* {
            (b) -- [fermion] (emVertex) -- [fermion, edge label=\large\( b \)] (weakInt) -- [fermion] (c),
            (v) -- [fermion] (weakInt),
            (weakInt) -- [fermion] (l),
            (emVertex) -- [boson] (photon),
            (d2)  -- [fermion] (d1)
        };
        \draw [decoration={brace}, decorate] (d1.south west) -- (b.north west)
            node [pos=0.5, left] {\Large \( \bar{B} \;  \)};
        \draw [decoration={brace}, decorate] (c.north east) -- (d2.south east)
            node [pos=0.5, right] {\Large \( \; D^{(*)} \)};
    \end{feynman}
\end{tikzpicture}}
    \caption{Feynman diagrams contributing to the $\bar{B} \rightarrow D^{(*)} \ell \bar{\nu}_\ell \gamma$ process. Diagrams with internal gluon lines and $q \bar{q}$ loops are omitted for illustrative simplicity. \textbf{Top Left}: $\mathcal{M}_\text{lepton}$, radiation from the external lepton. \textbf{Top Right}: $\mathcal{M}_\text{heavy}^c$, radiation from the heavy $c$ quark. \textbf{Bottom Left}: $\mathcal{M}_\text{light}$, radiation from the light degrees of freedom. \textbf{Bottom Right}: $\mathcal{M}_\text{heavy}^b$, radiation from the heavy $b$ quark.}
    \label{fig:feynman_diagrams}
\end{figure*}

Since $\bar{B} \rightarrow D^{(*)} \ell \bar{\nu}_\ell \gamma$ is a radiative process all matrix elements will have a photon polarization vector, $\varepsilon^\mu$ appended. We factor this out immediately, defining $\mathcal{M} \equiv \varepsilon_\mu^* \mathcal{M}^\mu$ where $\mathcal{M}$ is the total matrix element. In the heavy quark limit there are three classes of Feynman diagrams that can contribute to this process, shown in Fig.~\ref{fig:feynman_diagrams}. The first, illustrated in the top left panel of Fig.~\ref{fig:feynman_diagrams}, is from radiation off the external charged lepton line, 
\begin{align}
    \frac{\mathcal{M}_\text{lepton}^\mu}{\sqrt{m_B m_{D^{(*)}}}} & = q_\ell \frac{G_F e V_{cb}}{\sqrt{2}} \frac{(F_V^\nu - F_A^\nu)}{2 p_\ell \cdot k} \bar{u} \gamma^\mu ( \slashed{p}_\ell + \slashed{k} + m_\ell) \gamma_\nu ( 1 - \gamma^5 ) v \,
    \label{eq:M_lepton}
\end{align}
where $q_\ell e$ is the electromagnetic charge of the lepton, $p_\ell$ is the lepton momentum, and $k$ is the photon momentum. We write this matrix element in terms of hadronic matrix elements, $F_\Gamma^\nu$, where $\Gamma = V$ implies $\Gamma^\nu = \gamma^\nu$ and $\Gamma = A$ implies $\Gamma^\nu = \gamma^\nu \gamma^5$.\footnote{We will adopt the standard~\cite{Manohar:2000dt, Isgur:1989vq} notation and normalization for the states in the heavy quark effective theory (HQET). Relativistic states are labelled with $p$, e.g., $| \bar{B}(p) \rangle$, and HQET states are labelled with the velocity, $v$, e.g. $| \bar{B}(v) \rangle$. The normalization of these states differs by a factor of $\sqrt{m}$: $| \bar{B}(p) \rangle = \sqrt{m_B} | \bar{B}(v) \rangle$. This explains the extra factor of $\sqrt{m_B m_{D^{(*)}}}$ in the matrix elements.} This is a textbook example of how HQS severely restricts the number of independent form factors. In the heavy quark limit $F_\Gamma^\nu$, is proportional to just one overall factor, $\xi(w)$, the Isgur-Wise function, which depends only on $w = v^{\prime} \cdot v$,
\begin{align}
    F_\Gamma^\nu = - \xi(w) \left\langle D^{(*)}(v') \left| \text{Tr}\left[ \bar{H}_{v'}^c \Gamma^\nu H_v^b \right] \right| \bar{B}(v) \right\rangle
\end{align}
as in Ref.~\cite{Manohar:2000dt}, where $H_{v}^Q$ is defined and the traces are explicitly evaluated.

We now turn to radiation off the light degrees of freedom, $\mathcal{M}_\text{light}^\mu$, e.g., the Feynman diagram in the bottom left panel of Fig.~\ref{fig:feynman_diagrams}. The total matrix element is,
\begin{align}
    \frac{\mathcal{M}_\text{light}^\mu}{\sqrt{m_B m_{D^{(*)}}}} & = \frac{G_F e V_{cb}}{\sqrt{2}} \left( T_V^{\mu \nu} - T_A^{\mu \nu} \right) \bar{u} \gamma_\nu (1 - \gamma^5) v \, , 
    \label{eq:M_light}
\end{align}
where
\begin{align}
T^{\mu \nu}_\Gamma = \left\langle D^{(*)}(v') \left| \text{Tr} \left[ X^\mu \bar{H}_{v'} \Gamma^\nu H_v \right] \right| \bar{B}(v) \right\rangle \label{eq:T_mu_nu_Gamma_HQET}
\end{align}
and $X^\mu$ describes the effects of the time-ordered product between the light degrees of freedom electromagnetic current and the heavy quarks' weak operator. Again HQS limits the general form of $T^{\mu \nu}_\Gamma$, where the most general form of $X^\mu$ is,
\begin{align}
    X^\mu = x^{(1)} v^\mu + x^{(2)} v^{\prime \mu} + x^{(3)} \gamma^\mu + i x^{(4)} \sigma^{\mu \nu} k_\nu + i x^{(5)} \epsilon^{\mu \nu \rho \lambda} v_\nu v'_\rho k_\lambda \gamma^5 + x^{(6)} (v^\mu + v'^\mu) \slashed{k} \,,
    \label{eq:X_form_factor}
\end{align}
$\sigma^{\mu \nu} \equiv (i/2) [ \gamma^\mu, \gamma^\nu]$ and $\epsilon^{0123} = 1$. We omitted terms $\propto k^\mu$, since they give vanishing contribution to the matrix element, as well as terms that do not transform as vectors under parity since QCD and QED are parity invariant theories. HQS forbids $X^\mu$ to be between $\bar{H}_{v'}$ and $H_v$, and therefore the most general form is given by Eq.~\eqref{eq:T_mu_nu_Gamma_HQET}.\footnote{While not immediately obvious, a term $\propto (v^\mu - v'^\mu) \slashed{k}$ in $X^\mu$ can in fact be reabsorbed into the other terms present in Eq.~\eqref{eq:X_form_factor}.} The $x^{(j)}$ form factors are scalar functions of $w$, $v\cdot k$, $v' \cdot k$. $X^\mu$ has mass dimension $-1$, therefore $x^{(1-3)}$ have mass dimensions $-1$ while $x^{(4-6)}$ have mass dimension $-2$.

The last class of Feynman diagrams is from radiation off the heavy quarks, shown in the right column of Fig.~\ref{fig:feynman_diagrams}. There are two contributions here due to radiation from the heavy quark in the initial and final states. The matrix element is given by
\begin{align}
    \frac{\mathcal{M}_\text{heavy}^\mu}{\sqrt{m_B m_{D^{(*)}}}} & = \frac{G_F e V_{cb}}{\sqrt{2}} \left( q_b v^\mu (G_{V, b}^\nu - G_{V, b}^\nu) + q_c v'^\mu (G_{V, c}^\nu - G_{A, c}^\nu)\right) \bar{u} \gamma_\nu (1 - \gamma^5) v \,,
    \label{eq:M_heavy}
\end{align}
where
\begin{align}
    G_{\Gamma, Q}^\nu = \left\langle D^{(*)}(v') \left| \text{Tr}\left[ Y_Q \bar{H}_{v'} \Gamma^\nu H_v \right] \right| \bar{B}(v) \right\rangle \, 
\end{align}
and $Y_Q$ describes the time ordered product between the number density of the heavy quarks, $\bar{Q} Q$, and the heavy quarks weak operator. The absence of a $\gamma^\mu$ factor from the heavy quark electromagnetic vertex is because we are working in the heavy quark limit which replaces $\slashed{D} \rightarrow v \cdot D$ when acting on the heavy quark fields. The most general form of the $Y_Q$'s allowed by HQS are
\begin{align}
    Y_Q = y_Q^{(1)} + y_Q^{(2)} \slashed{k}\, , 
    \label{eq:Y_form_factor}
\end{align}
and $y_Q^{(1)}$ has mass dimension $-1$, $y_Q^{(2)}$ has mass dimension $-2$, and both are functions of $w, v \cdot k$ and $v' \cdot k$. Summing the matrix elements in Eqs.~(\ref{eq:M_lepton},~\ref{eq:M_light}, \ref{eq:M_heavy}) gives the total matrix element,
\begin{align}
    \frac{\mathcal{M}^\mu}{\sqrt{m_B m_{D^{(*)}}}} = \frac{G_F e V_{cb}}{\sqrt{2}} \bigg( & \left( K_V^{\mu \nu} - K_A^{\mu \nu} \right) \bar{u} \gamma_\nu (1 - \gamma^5) v \nonumber \\
    & + (q_b - q_c) \frac{(F_V^\nu - F_A^\nu)}{2 p_\ell \cdot k} \bar{u} \gamma^\mu (\slashed{p}_\ell + \slashed{k} + m_\ell) \gamma_\nu (1- \gamma^5) v \bigg) \, ,
    \label{eq:M_total}
\end{align}
where 
\begin{align}
    K_\Gamma^{\mu \nu} & \equiv T_\Gamma^{\mu \nu} + q_b v^\mu G_{\Gamma, b}^\nu + q_c v'^\mu G_{\Gamma, c}^\nu \, ,
\end{align}
and we have replaced $q_\ell$ with $q_b - q_c$ which will prove convenient in Sec.~\ref{subsec:ward_identities}.

\section{Form Factors}
\label{sec:form_factors}

In Sec.~\ref{sec:matrix_element} we showed that HQS limits the number of form factors, for both the $\bar{B} \rightarrow D \ell \bar{\nu}_\ell \gamma$ and $\bar{B} \rightarrow D^* \ell \bar{\nu}_\ell \gamma$ processes, to ten, $x^{(1-6)}, y_{b, c}^{(1, 2)}$. These ten form factors discussed so far are a superset of the number of physical form factors since we have not accounted for the Ward identity (WI). Next we show that the WI limits the number of new, independent form factors to just four. This is significant because, in general, even after applying the WI, the number of form factors is much larger. It has been shown in Ref.~\cite{Gasser:2004ds, Cirigliano:2005ms} that there are eight form factors for the $\bar{B} \rightarrow D \ell \bar{\nu}_\ell \gamma$ process, and in Appendix~\ref{app:expanded_form_factors} we show that the general number of form factors for the $\bar{B} \rightarrow D^* \ell \bar{\nu}_\ell \gamma$ process is 24. 

\subsection{Ward Identities}
\label{subsec:ward_identities}

The WI~\cite{Ward:1950xp} states that $k_\mu \mathcal{M}^\mu$ must vanish. This will highly constrain the form factors in Eqs.~(\ref{eq:X_form_factor},~\ref{eq:Y_form_factor}). While the WI is seemingly only one equation, different terms have to cancel each other individually. For example the $\Gamma = V, A$ terms have different parity, and therefore must cancel individually (as in Refs.~\cite{Bernlochner:2010yd, Gasser:2004ds, Cirigliano:2005ms, Becirevic:2009fy}) giving two equations,
\begin{align}
    & k_\mu K^{\mu \nu}_\Gamma + (q_b - q_c) F^\nu_\Gamma = k_\mu T^{\mu \nu}_\Gamma + q_b (k \cdot v) G_{\Gamma, b}^\nu + q_c (k \cdot v') G_{\Gamma, c}^\nu + (q_b - q_c) F_\Gamma^\nu = 0
\end{align}
for each $\Gamma \in \{ V, A \}$. Moreover, the electromagnetic charges, $q_i e$, give us another way to split the terms. Only $T^{\mu \nu}_\Gamma$ can depend on the charge of the light quark it was radiated from (either sea or valence), and therefore $k_\mu T^{\mu \nu}_\Gamma$ must vanish independently since there is no way to get a cancellation from the other terms which are independent of the light quark charges. Similar reasoning holds for the terms depending on $q_b, q_c$; all of the electromagnetic charge dependence has been factored out, and therefore $G^\nu_\Gamma$ and $F^\nu_\Gamma$ are independent of $q_b, q_c$. The terms proportional to $q_b, q_c$ must then cancel individually. The WI becomes six equations,
\begin{align}
    & k_\mu T^{\mu \nu}_\Gamma = 0 \label{eq:WI_1} \\
    & (k \cdot v) G_{\Gamma, b}^\nu + F_\Gamma^\nu = 0 \label{eq:WI_2} \\
    & (k \cdot v') G_{\Gamma, c}^\nu - F_\Gamma^\nu = 0 \label{eq:WI_3}
\end{align}
where, again, there is a set of equations for $\Gamma \in \{ V, A \}$.

We begin by simplifying the $\Gamma = A$ case for Eqs.~(\ref{eq:WI_1}-\ref{eq:WI_3}) for the process $\bar{B} \rightarrow D \ell \bar{\nu}_\ell \gamma$. The left side of Eqs. (\ref{eq:WI_2},~\ref{eq:WI_3}) are proportional to a single form factor, and therefore these must be zero. Specifically,
\begin{align}
    y^{(2)}_Q & = 0
\end{align}
for $Q \in \{ b, c \}$. Simplifying Eq.~\eqref{eq:WI_1} gives
\begin{align}
    -i \epsilon^{\mu \nu \rho \sigma} k_\nu v_\rho v'_\sigma \left( k \cdot (v + v') x^{(6)} + x^{(3)} \right) = 0
\end{align}
and therefore relates $x^{(3)}$ to $x^{(6)}$,
\begin{align}
    x^{(3)} = - k \cdot (v + v') \, x^{(6)} \, .
\end{align}

Next, we consider the case of $\Gamma = V$, again for the $B \rightarrow D \ell \bar{\nu}_\ell \gamma$ scenario, and Eqs.~(\ref{eq:WI_1}-\ref{eq:WI_3}) become
\begin{align}
    & -(v^\nu + v'^\nu) ( x^{(1)} (k \cdot v) + x^{(2)} (k \cdot v') ) = 0 \\
    & (v^\nu + v'^\nu) ( \xi - y_b^{(1)} (k \cdot v) ) = 0 \\
    & -(v^\nu + v'^\nu) ( \xi + y_c^{(1)} (k \cdot v') ) = 0
\end{align}
respectively, and therefore
\begin{align}
    x^{(2)} & = - \frac{k \cdot v}{k \cdot v'} x^{(1)} \nonumber \\
    y_b^{(1)} & = \frac{\xi}{k \cdot v} \nonumber \\
    y_c^{(1)} & = -\frac{\xi}{k \cdot v'} \, .
\end{align}
The WI has taken us from ten independent form factors, $x^{(1-6)}$, $y_{b, c}^{(1-2)}$, to just four. We define the independent ones as $\zeta_{1-4}$,
\begin{align}
    \zeta_1 & \equiv (k \cdot v) x^{(1)}\nonumber  \\
    \zeta_2 & \equiv x^{(4)}\nonumber  \\
    \zeta_3 & \equiv x^{(5)}\nonumber  \\
    \zeta_4 & \equiv x^{(6)}
\end{align}
(the motivation for the extra factor of $v \cdot k$ will become clear in Sec.~\ref{sec:soft_limit}). $\zeta_1$ is dimensionless while $\zeta_{2-4}$ have dimension of $\textrm{mass}^{-2}$. Written in terms of the five independent form factors (including the Isgur-Wise function) the $x, y$ form factors are
\begin{align}
    & x^{(1)} = \frac{\zeta_1}{(k \cdot v)} \quad , \quad x^{(2)} = -\frac{\zeta_1}{(k \cdot v')}\nonumber  \\
    & y_b^{(1)} = \frac{\xi}{(k \cdot v)} \quad , \quad y_c^{(1)} = -\frac{\xi}{(k \cdot v')}\nonumber  \\
    & x^{(3)} = - k \cdot (v + v') \zeta_4\nonumber  \\
    & y_b^{(2)} = y_c^{(2)} = 0\nonumber  \\
    & x^{(4)} = \zeta_2 \quad , \quad x^{(5)} = \zeta_3 \quad , \quad x^{(6)} = \zeta_4
\end{align}
One can check that with these conditions the WI for the $\bar{B} \rightarrow D^* \ell \bar{\nu}_\ell \gamma$ scenario are automatically satisfied. Using these relations, for the $\bar{B} \rightarrow D \ell \bar{\nu}_\ell \gamma$ case, the tensor $T^{\mu \nu}_\Gamma$ simplifies to  
\begin{align}
    T^{\mu \nu}_V & = \zeta_1 \left( \frac{v'^\mu}{k \cdot v'} - \frac{v^\mu}{k \cdot v} \right) (v^\nu + v'^\nu) + \zeta_2 \, \left( g^{\mu \nu} k \cdot (v' - v) + (v^\mu - v'^\mu)k^\nu \right) \nonumber \nonumber \\
    & + \zeta_4 \left( (w - 1) \left( g^{\mu \nu} (k \cdot (v + v')) - k^\nu (v^\mu + v'^\mu) \right)  + (v'^\nu - v^\nu) \left(  (k \cdot v) v'^\mu - (k \cdot v') v^\mu \right)\right) \label{eq:T_V_B_D}\\
    iT^{\mu \nu}_A & = \zeta_2 \, \epsilon^{\mu \nu \rho \sigma} k_\rho (v_\sigma + v'_\sigma) + \zeta_3 \epsilon^{\mu \rho \lambda \sigma} k_\rho v_\lambda v'_\sigma (v'^\nu - v^\nu) + \zeta_4 \bigg( (v^\mu + v'^\mu) \epsilon^{\nu \rho \sigma \lambda} k_\rho v_\sigma v'_\lambda \nonumber \\
    & + k \cdot (v + v') \epsilon^{\mu \nu \rho \sigma} v_\rho v'_\sigma \bigg) \label{eq:T_A_B_D}
\end{align}
and $G_{\Gamma, i}^\nu$ are reduced to
\begin{align}
    G^\nu_{V, b} & = - \frac{\xi}{k \cdot v} (v^\nu + v'^\nu) \label{eq:G_V_b_B_D} \\ 
    G^\nu_{A, b} & = 0 \\
    G^\nu_{V, c} & = \frac{\xi}{k \cdot v'} (v^\nu + v'^\nu) \label{eq:G_V_c_B_D} \\
    G^\nu_{A, c} & = 0 \, . 
\end{align}
For the $\bar{B} \rightarrow D^{*} \ell \bar{\nu}_\ell \gamma$ process, the $T^{*\,\mu \nu}_\Gamma$ read
\begin{align}
    T^{*\, \mu \nu}_V & = \zeta_1 \, {\epsilon^{\nu \rho \sigma}}_{\lambda} v_\rho v'_\sigma \varepsilon^{*\,\lambda}_{D^*} \left( \frac{v'^\mu}{k \cdot v'} - \frac{v^\mu}{k \cdot v} \right) \nonumber\\
    & + \zeta_2 \, \bigg(-\varepsilon_{D^*}^{*\,\mu} \epsilon^{\nu \rho \sigma \lambda} k_\rho v_\sigma v'_\lambda + v^\nu {\epsilon^{\mu \rho \sigma}}_{\lambda} k_\rho v'_\sigma \varepsilon_{D^*}^{*\,\lambda} + v'^\nu {\epsilon^{\mu \rho \sigma}}_{\lambda} k_\rho v_\sigma \varepsilon_{D^*}^{*\,\lambda} \nonumber \\
    & - (k \cdot \varepsilon_{D^*}^*) \epsilon^{\mu \nu \rho \sigma} v_\rho v'_\sigma + (w - 1) {\epsilon^{\mu \nu \rho}}_{\sigma} k_\rho \varepsilon_{D^*}^{*\,\sigma} \bigg) + \zeta_3 \epsilon^{\mu \rho \sigma \lambda} k_\rho v_\sigma v'_\lambda \bigg( (w - 1) \varepsilon_{D^*}^{*\, \nu}  - v'^\nu (v \cdot \varepsilon_{D^*}^*)\bigg) \nonumber \\
    & + \zeta_4 \left( (v \cdot \varepsilon_{D^*}^*) \epsilon^{\mu \nu \rho \sigma} k_\rho (v - v')_\sigma  - (v^\nu + v'^\nu) {\epsilon^{\mu \rho \sigma}}_{\lambda} k_\rho  (v_\sigma - v'_\sigma)\varepsilon_{D^*}^{*\, \lambda} \right) \nonumber
\end{align}
\begin{align}
    i T^{*\, \mu \nu}_A & = \zeta_1 \left( \frac{v'^\mu}{k \cdot v'} - \frac{v^\mu}{k \cdot v} \right) \left((w + 1) \varepsilon_{D^*}^{*\,\nu} - v'^\nu (v \cdot \varepsilon_{D^*}^*) \right) \nonumber \\
    & - \zeta_2 \Bigg( k^{\nu} \left((w+1) \varepsilon_{D^*}^{*\,\mu} - v'^{\mu}(v \cdot \varepsilon_{D^*}^*)\right) - g^{\mu \nu} \left( (w + 1) (k \cdot \varepsilon_{D^*}^*) - (k \cdot v') (v \cdot \varepsilon_{D^*}^*) \right) \nonumber \\
    & + v^{\prime \{\mu } v^{\nu \} } (k \cdot \varepsilon_{D^*}^*) - v^{ \{\mu } \varepsilon_{D^*}^{*\, \nu \} } (k \cdot v') + v^{\prime \, [ \mu } \varepsilon_{D^*}^{*\, \nu ] } (k \cdot v) \Bigg) \nonumber \\
    & + \zeta_3 \bigg( (1 - w^2) \left( (k \cdot \varepsilon_{D^*}^*) g^{\mu \nu} - \varepsilon_{D^*}^{*\,\mu} k^\nu \right) - (v \cdot \varepsilon_{D^*}) \left( k \cdot (v - w v') g^{\mu \nu} - (v^\mu - w v'^\mu) k^\nu \right)  \nonumber \\
    & + v^\nu \left( k \cdot (v- wv') \varepsilon_{D^*}^{*\,\mu} - (k \cdot \varepsilon_{D^*}^*) (v^\mu - w v'^\mu)\right) + v'^\nu (v \cdot \varepsilon_{D^*}^*) \left( (k \cdot v) v'^\mu - (k \cdot v') v^\mu \right) \nonumber \\
    & + v'^\nu \left( k \cdot (v' - wv) \varepsilon_{D^*}^{*\,\mu} - (k \cdot \varepsilon_{D^*}^*) (v'^\mu - wv^\mu ) \right) \bigg) \nonumber \\
    & + \zeta_4 \bigg( (v \cdot \varepsilon_{D^*}^*) \left( g^{\mu \nu} (k \cdot (v + v')) - (v^\mu + v'^\mu) k^\nu \right) \nonumber \\
    & + (v'^\nu - v^\nu) \left( \varepsilon^{*\,\mu}_{D^*} ( k \cdot (v + v') ) - (k \cdot \varepsilon_{D^*}^*) (v^\mu + v'^\mu) \right) \bigg)
\end{align}
and $G_{\Gamma, i}^{*\,\nu}$ are reduced to
\begin{align}
    G^{*\, \nu}_{V, b} & = - \frac{\xi}{k \cdot v} {\epsilon^{\nu \rho \sigma}}_{\lambda} v_\rho v'_\sigma \varepsilon_{D^*}^{*\,\lambda} \nonumber \\ 
    i G^{*\, \nu}_{A, b} & = -\frac{\xi}{k \cdot v} \left( (w + 1) \varepsilon_{D^*}^{*\, \nu} - v'^\nu (v \cdot \varepsilon_{D^*}^* )\right) \nonumber \\
    G^{*\, \nu}_{V, c} & = \frac{\xi}{k \cdot v'} {\epsilon^{\nu \rho \sigma}}_{\lambda} v_\rho v'_\sigma \varepsilon_{D^*}^{*\,\lambda} \nonumber \\
    iG^{*\, \nu}_{A, c} & = \frac{\xi}{k \cdot v'} \left( (w + 1) \varepsilon_{D^*}^{*\, \nu} - v'^\nu (v \cdot \varepsilon_{D^*}^* )\right)
\end{align}
where $a^{ [ \mu} b^{ \nu ]} \equiv a^\mu b^\nu - b^\mu a^\nu$ (and is straightforwardly generalized for the anti-commutator) and $\varepsilon_{D^*}^\mu$ is the $D^*$ polarization vector.

\section{Soft Limit}
\label{sec:soft_limit}

\begin{figure}[t]
    \centering
    \begin{tikzpicture}
    \begin{feynman}
        \vertex at (0, 0) (b) {\Large \( \bar{B}_a \)};
        \vertex at (6cm, 0) (c) {\Large \( D_a^{(*)} \)};
        \vertex[blob] at ($(b)!0.67!(c)$) (weakInt) {};
        \vertex at ($(weakInt)!0.5!(c) + (0, 1.5cm)$) (lvVertex);
        \vertex[above=1cm of c] (v) {\large \( \bar{\nu}_\ell \)};
        \vertex[above=2cm of c] (l) {\large \( \ell \)};
        \vertex at ($(b)!0.5!(weakInt)$) (emVertex);
        \vertex[below=1cm of d2] (photon) {\large \( \gamma \)};
        \diagram* {
            (b) -- [fermion] (emVertex) -- [fermion, edge label=\Large\( \bar{B}_a \)] (weakInt) -- [fermion] (c),
            (v) -- [fermion] (weakInt),
            (weakInt) -- [fermion] (l),
            (emVertex) -- [boson] (photon)
        };
    \end{feynman}
\end{tikzpicture}
    \begin{tikzpicture}
    \begin{feynman}
        \vertex at (0, 0) (b) {\Large \( \bar{B}_a \)};
        \vertex at (6cm, 0) (c) {\Large \( D_a^{(*)} \)};
        \vertex at ($(b)!0.67!(c)$) (emVertex);
        \vertex at ($(emVertex)!0.5!(c) + (0, 1.5cm)$) (lvVertex);
        \vertex[above=1cm of c] (v) {\large \( \bar{\nu}_\ell \)};
        \vertex[above=2cm of c] (l) {\large \( \ell \)};
        \vertex[blob] at ($(b)!0.5!(emVertex)$) (weakInt) {};
        \vertex[below=1cm of d2] (photon) {\large \( \gamma \)};
        \diagram* {
            (v) -- [fermion] (weakInt) -- [fermion, edge label'=\Large\( D_a^{(*)} \)] (emVertex) -- [fermion] (c),
            (b) -- [fermion] (weakInt),
            (weakInt) -- [fermion] (l),
            (emVertex) -- [boson] (photon)
        };
    \end{feynman}
\end{tikzpicture}

    \caption{In the soft limit, $k \ll \Lambda_\text{QCD}$, the leading order contributions to the radiative semileptonic $\bar{B}_a \rightarrow D_a^{(*)} \ell \bar{\nu}_\ell \gamma$ processes ($\mathcal{M} \sim k^{-1}$)  originate from these Feynman diagrams in the heavy hadron effective theory.}
    \label{fig:leading_soft_feynman_diagrams}
\end{figure}
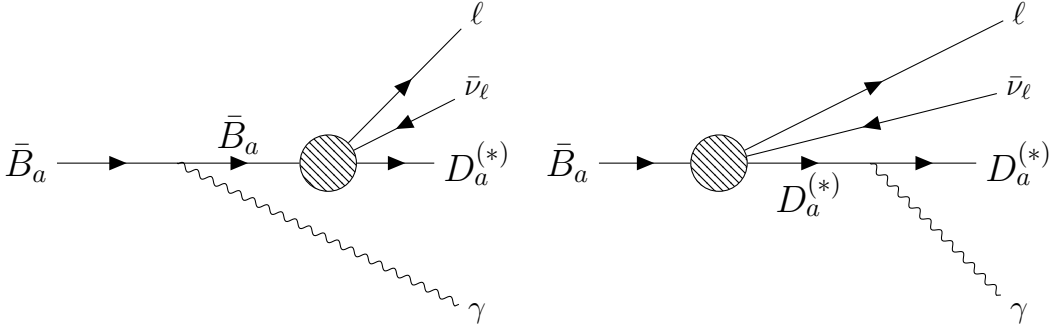

Further insight on the structure of the new form factors $\zeta_{1-4}$ derived in Sec.~\ref{subsec:ward_identities} can be gained by studying the small $k$ (soft) limit, $k \ll \Lambda_\text{QCD}$, where leading and sub-leading soft theorems~\cite{Low:1958sn,Weinberg:1965nx,Jackiw:1968zza} constrain the $k^{-1}$ and $k^0$ behavior of the $\zeta_i$'s, relating the $\bar B \rightarrow D^{(*)} \ell \bar{\nu}_\ell \gamma$ amplitudes to the amplitudes without photon emission, $\bar B \rightarrow D^{(*)} \ell \bar{\nu}_\ell$.

The soft limit is a strict subset of the $m_Q \rightarrow \infty$ limit, and can be described in terms of hadron fields. This limit is accessed by matching heavy quark effective theory onto a heavy hadron effective theory containing only the lowest multiplets. The effect of higher multiplets are encoded in higher derivative operators. We make the $SU(3)_V$ transformation properties of the heavy hadron fields explicit in this section.

To understand the leading, and sub-leading order, behavior we expand the matrix element in a power series in $k$: $\mathcal{M} = \mathcal{M}_{-1} + \mathcal{M}_0 + \mathcal{O}(k)$, where $\mathcal{M}_i \sim \mathcal{O}(k^i)$. In Sec.~\ref{subsec:leading_soft} we compute $\mathcal{M}_{-1}$ exactly, and then in Sec.~\ref{subsec:subleading_soft} we discuss the calculation of $\mathcal{M}_0$ and compute explicitly the contribution to it from the magnetic dipole operator. We find that in both cases the $\zeta_i$'s can be written in terms of the Isgur-Wise function.

\subsection{Leading Soft Behavior}
\label{subsec:leading_soft}

To explore the $k^{-1}$ term in the amplitude it is sufficient to study the leading order heavy hadron Lagrangian. At this order, the interaction with the photon is described by the mimimal coupling of the heavy hadron field to the electromagnetic field,
\begin{align}
    \left( D_\mu H \right)_a & = \partial_\mu H_a - i e A_\mu Q_{ab} H^b \label{eq:covariant_derivative}\\
    \mathbf{Q} & = \text{diag} \left( q_Q - q_u, q_Q - q_d, q_Q - q_s \right) \nonumber
\end{align}
where $a, b$ are $SU(3)_V$ flavor indices and $q_u e$, $q_d e$, $q_s e$ are the electromagnetic charges of the up, down, and strange quarks, respectively. The expression $e(q_Q -q_a)$ is just the electric charge of the heavy meson written in terms of the electric charge of the heavy quark, $q_Q e$, and the light valence quark, $q_a e$, as it will be convenient in the matching to the heavy quark limit expressions. The interaction Lagrangian is then
\begin{align}
    \mathcal{L}_\text{int} = e v^\mu A_\mu \text{Tr}\left[ \bar{H} \cdot \mathbf{Q} \cdot H \right]
    \label{eq:chiral_limit_L_int}
\end{align}
With the interaction Lagrangian, along with the hadron field propagator, $i/(2k \cdot v)$, we can compute the matrix element for $\bar B_a \rightarrow D_a^{(*)} \ell \bar{\nu}_\ell \gamma$, shown in Fig.~\ref{fig:leading_soft_feynman_diagrams}. We will use the convention $\bar B_a$, $D^{(*)}_a$ with $a = u,d,s$ based on the valence quark flavor to describe different processes. The matrix element, $\mathcal{M}_{-1}$, with just the leading order contribution from radiation off the hadron fields, is
\begin{align}
    \frac{\mathcal{M}^\mu_{-1}}{\sqrt{m_B m_{D^{(*)}}}} = \frac{G_F e V_{cb}}{\sqrt{2}} \left[ \left( \frac{(q_a - q_b) v^\mu}{k \cdot v} + \frac{(q_c - q_a) v'^\mu}{k \cdot v'} \right) \left( F_V^\nu - F_A^\nu \right) \right] \bar{u} \gamma_\nu (1 - \gamma^5) v + \frac{\mathcal{M}^\mu_{-1,\text{lepton}}}{\sqrt{m_B m_{D^{(*)}}}} \, .
\end{align}
where $\mathcal{M}_\text{lepton}^\mu$ here is the same as in Sec.~\ref{sec:matrix_element}, and the other two terms describe the photon emission from the $\bar B_a$ and the $D^{(*)}_a$ lines, and are collectively $\mathcal{M}_{-1}$. Matching this on to the matrix element found previously, Eq.~\eqref{eq:M_total}, expanded in $k$ at this order, it is easy to confirm that the terms $\propto q_b, q_c$ are automatically matched to Eqs.~(\ref{eq:G_V_b_B_D},~\ref{eq:G_V_c_B_D}), while the $q_a$ dependence must be contained in $T^{\mu \nu}$. Considering the $\bar{B}_a \rightarrow D_a \ell \bar{\nu}_\ell \gamma$ case and the terms $\propto q_a$,
\begin{align}
    q_a \xi \, \left( \frac{v^\mu}{k \cdot v} - \frac{v'^\mu}{k \cdot v'}\right) (v^\nu + v'^\nu) = T^{\mu \nu}_V
\end{align}
where $T^{\mu \nu}_V$ is given in Eq.~(\ref{eq:T_V_B_D}). Therefore we see that
\begin{align}
    \zeta_1 \rightarrow -q_a \xi + \mathcal{O} \left ( \frac{k}{\Lambda_\text{QCD}} \right) \, ,
\end{align}
which justifies the extra factor of $v \cdot k$ in the redefinition of $x^{(1)} \rightarrow \zeta_1$. We also learn that $\zeta_{2-4}$ are unconstrained and can begin at order $\mathcal{O}(1/v \cdot k, 1/ v' \cdot k)$, as we will see in the next section. To connect to the literature~\cite{Gasser:2004ds,Becirevic:2009fy,Bernlochner:2010yd}, the emissions from the lepton, the heavy quarks and the leading contribution to $\zeta_1$ correspond to what is conventionally called internal bremsstrahlung (IB) while $\mathcal{O}(k)$ contributions to $\zeta_1$, and $\zeta_{2-4}$, parameterize the structure dependent (SD) contributions.  

\subsection{Sub-Leading Soft Behavior}
\label{subsec:subleading_soft}

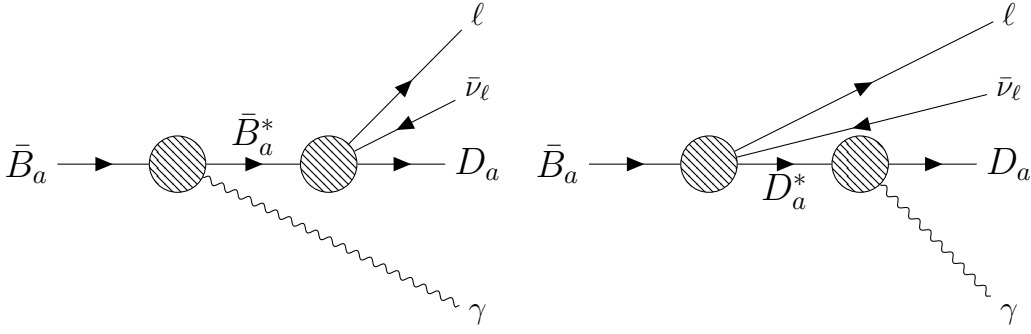
\begin{figure}
    \centering
    \begin{tikzpicture}
    \begin{feynman}
        \vertex at (0, 0) (b) {\Large \( \bar{B}_a \)};
        \vertex at (6cm, 0) (c) {\Large \( D_a \)};
        \vertex[blob] at ($(b)!0.67!(c)$) (weakInt) {};
        \vertex at ($(weakInt)!0.5!(c) + (0, 1.5cm)$) (lvVertex);
        \vertex[above=1cm of c] (v) {\large \( \bar{\nu}_\ell \)};
        \vertex[above=2cm of c] (l) {\large \( \ell \)};
        \vertex[blob] at ($(b)!0.5!(weakInt)$) (emVertex) {};
        \vertex[below=1cm of d2] (photon) {\large \( \gamma \)};
        \diagram* {
            (b) -- [fermion] (emVertex) -- [fermion, edge label=\Large\( \bar{B}_a^* \)] (weakInt) -- [fermion] (c),
            (v) -- [fermion] (weakInt),
            (weakInt) -- [fermion] (l),
            (emVertex) -- [boson] (photon)
        };
    \end{feynman}
\end{tikzpicture}
    \begin{tikzpicture}
    \begin{feynman}
        \vertex at (0, 0) (b) {\Large \( \bar{B}_a \)};
        \vertex at (6cm, 0) (c) {\Large \( D_a \)};
        \vertex[blob] at ($(b)!0.67!(c)$) (emVertex) {};
        \vertex at ($(emVertex)!0.5!(c) + (0, 1.5cm)$) (lvVertex);
        \vertex[above=1cm of c] (v) {\large \( \bar{\nu}_\ell \)};
        \vertex[above=2cm of c] (l) {\large \( \ell \)};
        \vertex[blob] at ($(b)!0.5!(emVertex)$) (weakInt) {};
        \vertex[below=1cm of d2] (photon) {\large \( \gamma \)};
        \diagram* {
            (v) -- [fermion] (weakInt) -- [fermion, edge label'=\Large\( D_a^* \)] (emVertex) -- [fermion] (c),
            (b) -- [fermion] (weakInt),
            (weakInt) -- [fermion] (l),
            (emVertex) -- [boson] (photon)
        };
    \end{feynman}
\end{tikzpicture}
    \caption{Feynman diagrams contributing to the sub-leading soft, $k \ll \Lambda_\text{QCD}$, behavior of the radiative semileptonic process $\bar{B}_a \rightarrow D_a \ell \bar{\nu}_\ell \gamma$ within a heavy hadron effective theory. Intermediate excited states, $\bar{B}_a^*$ and $D_a^*$, arise from the magnetic dipole term in the Lagrangian, Eq.~\eqref{eq:L_dipole}.}
    \label{fig:sub_leading_soft_feynman_diagrams}
\end{figure}

To understand the sub-leading soft behavior of $\zeta_{1-4}$ we need to study the amplitude at order $\mathcal{O}(k^0)$ and consider higher order operators in the heavy hadron Lagrangian. These are effective operators with one extra covariant derivative on a hadron line, including the magnetic dipole operator,\footnote{There is also the operator $\text{Tr}\left[ \bar H D^2 H \right] $, but it does not contribute to the on-shell matrix element.}
\begin{align}
    \mathcal{L}_\text{dipole} & = \frac {i e }{2} F_{\mu\nu} \text{Tr}\left[ \sigma^{\mu\nu} \bar H \cdot \boldsymbol{\mu} \cdot H\right] \label{eq:L_dipole}\\
    \boldsymbol{\mu} & = \text{diag}(\mu_u, \mu_d, \mu_s) \, ,
\end{align}
which induces a transition between a $D_a$ ($\bar{B}_a$) and $D_a^*$ ($\bar{B}_a^*$). At this order there are also corrections to the weak current operator,
\begin{align}
    \mathcal{L}_\text{current} & = a_{1, c}(w) \, v^\mu \text{Tr}\left[(D_\mu \bar H_c^{(v')}) \Gamma H_b^{(v)}\right]  + a_{2,c}(w) \, \text{Tr}\left[(\slashed D \bar H_c^{(v')}) \Gamma H_b^{(v)}\right]\nonumber \\
      & + a_{1,b} (w) \, v'^\mu \text{Tr}\left[\bar H_c^{(v')} \Gamma D_\mu H_b^{(v)}\right] + a_{2,b} (w) \, \text{Tr}\left[\gamma^\mu \bar H_c^{(v')} \Gamma D_\mu H_b^{(v)}\right]
\end{align}
where $D_\mu$ is the covariant derivative defined in Eq.~\eqref{eq:covariant_derivative}. The effect of the current corrections vanishes in the zero recoil limit, $v=v'$, as they can be eliminated using the equations of motion. The coefficient of the magnetic dipole operator can be extracted from the $D^* \rightarrow D \gamma$ rate~\cite{Amundson:1992yp} and has been also computed on the lattice~\cite{Becirevic:2009xp}, while the coefficients $a_{1-2,b-c}$ of the current corrections are currently unknown. The $\mathcal{O}(k^0)$ matrix element reads
\begin{equation}
    \frac{\mathcal{M}^\mu_{0}}{\sqrt{m_{B_a} m_{D^{(*)}_a}}} = \frac{G_F e V_{cb}}{\sqrt{2}} \left(\Delta T^{\mu\nu}_{V} -\Delta T^{\mu\nu}_{A} \right) \bar u \gamma_\nu (1-\gamma^5) v + \frac{\mathcal{M}^\mu_{0,\text{lepton}}}{\sqrt{m_{B_a} m_{D^{(*)}_a}}}
\end{equation}
with $\Delta T_{V,A}$ receiving contribution both from the dipole and the current operators above. To investigate their structure we will focus on the magnetic dipole contribution to the $\bar B_a \rightarrow D_a \ell \bar{\nu}_\ell \gamma$ amplitude~\cite{Becirevic:2009fy}; the $\bar B_a \rightarrow D^*_a \ell \bar{\nu}_\ell \gamma$ process will be parameterically similar. There are two diagrams contributing, one where the photon is emitted from the $\bar B_a$ line (transitioning to a $\bar B^*_a$) and one from the $D_a$ (to $D^*_a$), shown in Fig.~\ref{fig:sub_leading_soft_feynman_diagrams}. Their sum read:
\begin{align}
   \Delta T^{\mu\nu}_{V,\text{dipole}} &= \frac{ \mu_a}{2} \xi(w) \bigg\{ g^{\mu\nu} \left( \frac{k \cdot v}{k \cdot v'} +  \frac{k \cdot v'}{k \cdot v} -2 w \right) + \left(\frac{v'^\mu}{k \cdot v'} - \frac{v^\mu}{k \cdot v}\right) k^\rho v^{[\nu} v'^{\rho]}  \nonumber \\
    & + k^\nu \left[\frac{v^\mu}{k \cdot v}\left(w - \frac{k \cdot v}{k \cdot v'}\right) + \frac{v'^\mu}{k \cdot v'} \left(w - \frac{k \cdot v'}{k\cdot v}\right) \right] \bigg\}\nonumber \\
    \Delta T^{\mu\nu}_{A,\text{dipole}} &= \frac{ \mu_a}{2} \xi(w) \bigg[ (1+w) \epsilon^{\mu\nu\rho\sigma}k_\rho \left(\frac{v_\sigma}{k \cdot v} - \frac{v'_\sigma}{k \cdot v'}\right) - \left(\frac{v'^\nu}{k \cdot v'} + \frac{v^\nu}{k \cdot v}\right) \epsilon^{\mu\rho\sigma\lambda}k_\rho v_\sigma v'_\lambda \bigg]
\end{align}

Upon matching these results to the $\zeta_{1-4}(w, v \cdot k, v' \cdot k)$ form factors one find that they receive contributions:
\begin{align}
     \zeta_1 & = -q_a \xi(w) + \frac{\mu_a}{4} \left(k \cdot v - k \cdot v'\right)\,\xi(w) + \ldots, \ \ & \zeta_2 &= -\frac{\mu_a}{4} (1+w) \left(\frac{1}{k \cdot v} - \frac{1}{k \cdot v'}\right)\,\xi(w) + \ldots, \nonumber \\
    \zeta_3 & =  -\frac{\mu_a}{4} \left(\frac{1}{k \cdot v} - \frac{1}{k \cdot v'}\right)\,\xi(w) + \ldots, \ \ &\zeta_4 &= \frac{\mu_a}{4} \left(\frac{1}{k \cdot v} + \frac{1}{k \cdot v'}\right)\,\xi(w) + \ldots.
\end{align}
where the ellipsis include both contributions from the current corrections at $\mathcal{O}(k^0)$ and terms higher order in $k$. As expected, the dipole contribution to $\zeta_1$ starts at $\mathcal{O}(k)$, while the contribution to $\zeta_{2-4}$ begin at $\mathcal{O}(k^{-1})$, and are all proportional to the leading order Isgur-Wise function.

The presence of $(k \cdot v / k\cdot v')$ and $(k \cdot v'/ k \cdot v)$ factors in the amplitude at $\mathcal{O}(k^0)$ is not unexpected. Similar ratios are known to appear in the sub-leading soft limit for effective theories with massless particles~\cite{Elvang:2016qvq}. The contributions from the current operators to $\Delta T_{V,A}$ will induce similar contributions to $\zeta_{1-4}$ and induce in the amplitude ratios of $v\cdot k$ and $v' \cdot k$.

\section{Discussion}
\label{sec:conclusion}

We have studied the implications of heavy quark symmetry (HQS) for radiative semileptonic $\bar B$ decays. This study is motivated by the increase in data that will be available from Belle II and LHCb. Radiative semileptonic $\bar B$ decays are an interesting process since they probe HQS in a different kinematic regime than non-radiative semileptonic processes. Moreover they are a background for other semileptonic $\bar B$ decay measurements, such as the extraction of $V_{cb}$ and of the $R(D^{(*)})$ ratios, when the radiated photon goes undetected. We restricted our attention to the kinematic region where radiating the final state photon does not change the heavy quark four velocity. In this region HQS implies a dramatic reduction in the number of independent scalar form factors characterizing the decay amplitudes.

Throughout this paper we have worked to lowest order in the electromagnetic coupling and neglected $1/m_{c,b}$ corrections. There are kinematic regions where this will not be adequate. 

For very soft photons there are infrared and collinear (with the electron) divergences that should be resummed, and one loop corrections that should be included. However, these effects are already well studied in the literature. NLO QED effects when the photon is soft have been computed in~\cite{Atwood:1989em, Bernlochner:2010yd, deBoer:2018ipi}, including resumming the Coulomb corrections, while leading logs are partially resummed in \texttt{PHOTOS}~\cite{Barberio:1990ms, Barberio:1993qi} (up to quadratic order) or fully via YFS~\cite{Yennie:1961ad} in other MonteCarlo event generators such as \texttt{Sherpa}~\cite{Schonherr:2008av}. Several numerical comparisons between the different predictions of these calculations have been performed both in the B-factory~\cite{Bernlochner:2010fc} and LHCb~\cite{Cali:2019nwp} kinematic regimes. 

It is therefore important to compare our calculation with the approximations currently used by LHCb and Belle II in simulating semileptonic $\bar B$ decays. Experiments are currently using \texttt{PHOTOS} to generate radiated photons in semileptonic $\bar B$ decays (with the LO process usually generated by \texttt{EvtGen})\footnote{The program \texttt{BLOR} which include several effects neglected in \texttt{PHOTOS} was also available during the past B factory runs~\cite{Bernlochner:2010yd}. As there are no current plans for Belle II to use it, we will not discuss it any further.} and also include photons from $B\rightarrow (D^* \rightarrow D \gamma) \ell \bar{\nu}_\ell$ as a separate process with the $D^*$ (quasi) on-shell (i.e., $| m_{D\gamma} - m_{D^*} | < \textrm{few} \times \Gamma_{D^*}$). \texttt{PHOTOS} generates (multi-)photon radiation starting from a non-radiative decay event in the leading logarithmic approximation, extrapolated over the whole kinematic space allowed for photon emission, and neglects the interference between amplitudes where the photon is radiated from different charged legs. In the language of this paper, it is equivalent to setting $\zeta_1(w, v \cdot k, v' \cdot k) = - q_a \xi(w)$, $\zeta_{2-4}(w, k \cdot v, k \cdot v')=0$, neglecting the $\slashed{k}$ term in $\mathcal{M}_\textrm{lepton}$ and neglecting the interference between $\mathcal{M}_\textrm{lepton}$ and $\mathcal{M}_\textrm{light} + \mathcal{M}_\textrm{heavy}$ in $|\mathcal{M}|^2$. Furthermore, since the photon radiated from a charged $D^{(*)}$ is produced by a deformation of the kinematic configuration of the non-radiative process, for these diagrams the leading order Isgur-Wise function is evaluated at a different recoil parameter point $\bar w = (m_B^2+m^2_{D^{(*)}} - (p_B-p_{D^{(*)}\gamma})^2)/(2 m_B m_{D^{(*)}})$. However the difference $\bar w - w$~is~$\mathcal{O}(v^{(\prime)} \cdot k / m_{B,D^{(*)}})$ and therefore higher order in $1/m_{b,c}$.

The novel element of this paper is a compact parameterization of contributions to the radiative semileptonic $\bar{B} \rightarrow D^{(*)} \ell \bar{\nu}_\ell \gamma$ process in the region where the photon probes the structure of the heavy mesons, in the form of the (currently unknown) functions $\zeta_{1-4}$.

The effect of $\zeta_{1-4}$ (and other effects missing in \texttt{PHOTOS}) can be included in current event generators via, e.g., event reweighing with tools such as \texttt{Hammer}. The relevant helicity amplitudes, together with a numerical study of the size of the effects that $\zeta_{1-4}$ may induce in semileptonic measurements, along the lines of~\cite{Cali:2019nwp}, will be presented elsewhere.

We have also studied the low photon energy, $k < \Lambda_\text{QCD}$, behavior of $\zeta_{1-4}$ and found that the leading order, in $k$ behavior is determined by the Isgur-Wise function. The sub-leading contributions can be captured by the magnetic moment, $\mu$, and four Wilson coefficients $a_{1-2,b-c}$ which are functions of the recoil parameter $w$. ($\mu$ is constrained by the $D^* \rightarrow D \gamma$ width.) In particular, the results of Sec.~\ref{sec:soft_limit}, together with the bookkeeping procedure used by experiments to account for photons originating from on-shell decays of $D^* \rightarrow D \gamma$ in separate event samples, imply that, in practice, the magnetic dipole contributions to the SD form factors are never dominant. Away from the $D^*$ peak in the $(D\gamma)$ mass distribution the current corrections are expected to contribute to the same level (or even be parametrically larger when the photon energy drops below the $D-D^*$ ($B-B^*$) mass gap as the dipole operators will be further suppressed by the mass gap). 

There are various avenues to further pursue the study initiated here. For example, translating these results to a form that can be directly used by the experiments, and quantitative estimation of the impact of $\zeta_{1-4}$ on the semileptonic analyses as already mentioned above. With enough data Belle II can tag the photon and constrain $\zeta_{1-4}$. In practice a suitable parameterization of the $\zeta$'s amenable for fitting should be found. 
Study of radiative semileptonic transition can also be performed for other heavy hadrons, such as $\Lambda_b \rightarrow \Lambda_c \ell \nu$.
There are also more interesting field theoretical questions such as how to describe harder photon radiation in heavy quark effective theory. If the photon is sufficiently hard, the heavy quark line will have two cusps, one due to the weak vertex and one due to photon emission. 

\begin{acknowledgments}
We would like to thank Florian Bernlochner for discussions and Zoltan Ligeti and Dean Robinson for comments on the paper. This work is supported by the U.S. Department of Energy, Office of Science, Office of High Energy Physics, under Award Number DE-SC0011632 and by the Walter Burke Institute for Theoretical Physics.
\end{acknowledgments}

\appendix
\section{General Form Factor Expressions}
\label{app:expanded_form_factors}
For $\bar B \rightarrow D \ell \bar{\nu}_\ell \gamma$ the most general parameterization of the form factors read~\cite{Gasser:2004ds}:
\begin{align}
K^{\alpha\beta}_{V,\textrm{IB}}  & = \frac{1}{2}\left(q_{D}\frac{ v'^{\alpha }}{k \cdot v'}  - q_{\bar B}\frac{v^{\alpha }}{  k \cdot v} \right) \left(f_+(w) \left( v ^{\beta }+ v'^{\beta }\right)  + f_-(w) \left( v ^{\beta }- v'^{\beta }\right)\right)\\
K^{\alpha,\beta}_{V,\textrm{SD}} & = V^D_{1} \left( k ^{\beta }  v ^{\alpha }-g^{\alpha \beta }  k \cdot v \right) 
 + V^D_{2} \left( k ^{\beta }  v'^{\alpha }-g^{\alpha \beta }  k \cdot v' \right) \nonumber \\ &
 + \left(V^D_{3}  v'^{\beta } + V^D_{4}  v ^{\beta }\right)\left( v ^{\alpha }  k \cdot v' - v'^{\alpha }  k \cdot v \right) \\ 
K^{\alpha\beta}_{A,\textrm{IB}}  & = 0\\
i K^{\alpha,\beta}_{A,\textrm{SD}} & = A^D_{1} \epsilon ^{\alpha \beta  k  v } 
 + A^D_{2} \epsilon ^{\alpha \beta  k  v' } 
 + A^D_{3}  v ^{\beta } \epsilon ^{\alpha  k  v  v' } 
 + A^D_{4}  v'^{\beta } \epsilon ^{\alpha  k  v  v' }
\end{align}
where we have used $v^\alpha \equiv p_{\bar B}^\alpha / m_{\bar B}$, $v'^\alpha \equiv p_{D}^\alpha / m_{D}$ and the notation for the Levi-Civita tensor $\epsilon^{\beta v v' k} = \epsilon^{\beta \gamma\nu\rho}v^\gamma v'^\nu k^\rho$.
In the region $k \cdot v^{'} < m_{(b,c)}$ explored in this work, HQS fixes the new form factors to be:
\begin{align}
    & V_1^D  = \zeta_2 -(w-1) \zeta_4, & V_2^D & = -\zeta_2 - (w-1) \zeta_4,   \nonumber \\ 
    & V_3^D = -\frac{\zeta_1^\textrm{SD}}{k\cdot v k\cdot v'} - \zeta_4,  & V_4^D & = -\frac{\zeta_1^\textrm{SD}}{k\cdot v k\cdot v'} + \zeta_4,  
\end{align}
\begin{align}
    & A_1^D  = \zeta_2 -(1+w) \zeta_4, & A_2^D & = \zeta_2 + (1+w) \zeta_4, \nonumber \\
    & A_3^D = - \zeta_3 + \zeta_4,  & A_4^D & = \zeta_3 + \zeta_4 \, ,
\end{align}
where we have defined $\zeta_1^{\textrm{SD}}$ to be $\zeta_1 - \zeta_1^\textrm{IB}$ with $\zeta_1^\textrm{IB} \equiv q_a [ (v' \cdot \varepsilon)/(v\cdot k) - (v \cdot \varepsilon)/(v' \cdot k)] \xi(w)$.

The general parameterization for scalar to vector radiative semileptonic transitions are not present in the literature and we have derived them here. There are 12 structure-dependent form factors for both the vector and axial amplitudes. Therefore, for $\bar B \rightarrow D^* \ell \bar{\nu}_\ell \gamma$ the most general parameterization of the form factors read:

\begin{align}
i K^{\alpha\beta}_{V,\textrm{IB}}  & = -\left(q_{D^*}\frac{ v'^{\alpha }}{k \cdot v'}  - q_{\bar B}\frac{v^{\alpha }}{  k \cdot v} \right) g(w) \epsilon ^{\beta  v  v'  \varepsilon_{D^*}^* } \\
i K^{\alpha\beta}_{V,\textrm{SD}}  & =  V^{D^*}_{1} \left( v ^{\alpha } \epsilon ^{\beta  k  v  \varepsilon_{D^*}^* }+ k \cdot v  \epsilon ^{\alpha \beta  v  \varepsilon_{D^*}^* }\right) 
 + V^{D^*}_{2}  v ^{\beta }  v \cdot  \varepsilon_{D^*}^*  \epsilon ^{\alpha  k  v  v' } \nonumber \\ &
 + V^{D^*}_{3} \left( \varepsilon_{D^*}^{* \alpha } \epsilon ^{\beta  k  v  v' }+ k \cdot  \varepsilon_{D^*}^*  \epsilon ^{\alpha \beta  v  v' }\right) 
 + V^{D^*}_{4} \left( v ^{\alpha }  k \cdot v' - v'^{\alpha }  k \cdot v \right) \epsilon ^{\beta  v  v'  \varepsilon_{D^*}^* } \nonumber \\ &
 + V^{D^*}_{5} \epsilon ^{\alpha \beta  k  \varepsilon_{D^*}^* } 
 + \left(V^{D^*}_{6}  \epsilon ^{\alpha \beta  k  v } + V^{D^*}_{7}  \epsilon ^{\alpha \beta  k  v' } \right) k \cdot  \varepsilon_{D^*}^*  \nonumber \\ &
 + \left(V^{D^*}_{8}  v ^{\beta } + V^{D^*}_{9}  v'^{\beta }\right)\epsilon ^{\alpha  k  v  \varepsilon_{D^*}^* } 
 + \left(V^{D^*}_{10}  v ^{\beta } + V^{D^*}_{11}  v'^{\beta }\right)\epsilon ^{\alpha  k  v'  \varepsilon_{D^*}^* }  \nonumber \\ &
 + V^{D^*}_{12}  v \cdot  \varepsilon_{D^*}^*  \epsilon ^{\alpha \beta  k  v' } \\
K^{\alpha\beta}_{A,\textrm{IB}}  & = \frac{1}{2}\left(q_{D^*}\frac{ v'^{\alpha }}{ k \cdot  v' } - q_{\bar B}\frac{ v ^{\alpha }}{ k \cdot  v }\right) \times \nonumber \\ &
\left(f(w)  \varepsilon_{D^*}^{* \beta } + a_+(w) \left( v ^{\beta }+ v'^{\beta }\right) v \cdot  \varepsilon_{D^*}^* + a_-(w) \left( v ^{\beta }- v'^{\beta }\right)  v \cdot  \varepsilon_{D^*}^* \right)\\
K^{\alpha,\beta}_{A,\textrm{SD}} & = A^{D^*}_{1}  \varepsilon_{D^*}^{* \beta } \left( v ^{\alpha }  k \cdot v' - v'^{\alpha }  k \cdot v \right) 
 + A^{D^*}_{2} \left(\bar{g}^{\alpha \beta }  k \cdot  \varepsilon_{D^*}^* - \varepsilon_{D^*}^{* \alpha }  k ^{\beta }\right) \nonumber \\ &
 + A^{D^*}_{3}  k ^{\beta } \left( v ^{\alpha }  k \cdot  \varepsilon_{D^*}^* - \varepsilon_{D^*}^{* \alpha }  k \cdot v \right)   
 + A^{D^*}_{4}  k ^{\beta } \left( v'^{\alpha }  k \cdot  \varepsilon_{D^*}^* - \varepsilon_{D^*}^{* \alpha }  k \cdot v' \right) \nonumber \\ &
 + A^{D^*}_{5}  v \cdot  \varepsilon_{D^*}^*  \left( k ^{\beta }  v ^{\alpha }-g^{\alpha \beta }  k \cdot v \right) 
 + A^{D^*}_{6}  v \cdot  \varepsilon_{D^*}^*  \left( k ^{\beta }  v'^{\alpha }-g^{\alpha \beta }  k \cdot v' \right)   \nonumber \\ &
 + A^{D^*}_{7}  v ^{\beta } \left( v ^{\alpha }  k \cdot  \varepsilon_{D^*}^* - \varepsilon_{D^*}^{* \alpha }  k \cdot v \right) 
 + A^{D^*}_{8}  v ^{\beta } \left( v'^{\alpha }  k \cdot  \varepsilon_{D^*}^* - \varepsilon_{D^*}^{* \alpha }  k \cdot v' \right) \nonumber \\ &
 + A^{D^*}_{9}  v ^{\beta }  v \cdot  \varepsilon_{D^*}^*  \left( v ^{\alpha }  k \cdot v' - v'^{\alpha }  k \cdot v \right)   
 + A^{D^*}_{10}  v'^{\beta } \left( v ^{\alpha }  k \cdot  \varepsilon_{D^*}^* - \varepsilon_{D^*}^{* \alpha }  k \cdot v \right) \nonumber \\ &
 + A^{D^*}_{11}  v'^{\beta } \left( v'^{\alpha }  k \cdot  \varepsilon_{D^*}^* - \varepsilon_{D^*}^{* \alpha }  k \cdot v' \right) 
 + A^{D^*}_{12}  v'^{\beta }  v \cdot  \varepsilon_{D^*}^*  \left( v ^{\alpha }  k \cdot v' - v'^{\alpha }  k \cdot v \right)
\end{align}
Matching the amplitudes studied in this work in the region $k \cdot v^{'} < m_{(b,c)}$ to the most general parameterization, one finds that HQS fixes the new form factors in $\bar B \rightarrow D^* \ell \bar{\nu}_\ell \gamma$ to be:
\begin{align}
    & V_1^{D*} = - \zeta_3 + \zeta_4, & V_3^{D*} & = \zeta_2 - (w-1) \zeta_3 \nonumber \\
    & V_4^{D^*} = \frac{\zeta_1^\textrm{SD}}{k\cdot v\, k\cdot v'} - \zeta_3, & V_5^{D^*} & =(w-1)\zeta_2 -w^2 \zeta_3 + \zeta_4 \nonumber \\
    &V_9^{D^*}= -\zeta_2 +w \zeta_3 - \zeta_4, & V_{10}^{D^*} & = -\zeta_2 +w \zeta_3 + \zeta_4, \nonumber \\
    &V_{11}^{D^*}= - \zeta_3 - \zeta_4, & V_{12}^{D^*} & = \zeta_3 - \zeta_4, \nonumber \\
    & V_{2}^{D^*}= V_{6}^{D^*}= V_{7}^{D^*}= V_{8}^{D^*}= 0, \ \ & &
\end{align}
\begin{align}
    & A_1^{D^*}  = - (w+1)\frac{\zeta_1^\textrm{SD}}{k\cdot v\, k\cdot v'} + \zeta_2, & A_2^{D^*} & = (1+w) \zeta_2 - (w^2-1) \zeta_3, \nonumber\\
    & A_5^{D^*} = \zeta_3 - \zeta_4, & A_6^{D^*} &= \zeta_2 - w \zeta_3 - \zeta_4,  \nonumber \\  
    & A_7^{D^*} = -\zeta_3 + \zeta_4, & A_8^{D^*} &= -\zeta_2 + w \zeta_3 + \zeta_4, \nonumber \\  
    & A_{10}^{D^*} = - \zeta_2 + w \zeta_3 - \zeta_4, & A_{11}^{D^*} &=  - \zeta_3 - \zeta_4, \nonumber \\
    & A_{12}^{D^*} = \frac{\zeta_1^\textrm{SD}}{k\cdot v\, k\cdot v'} - \zeta_3 , & A_{3}^{D^*} & = A_{4}^{D^*} = A_{9}^{D^*} = 0 \, .
\end{align}

\section{Matrix Elements For Other Dirac Structures}
\label{app:other_matrix_elements}
Here we provide the matrix elements $T^{(*)\,\mu}_\Gamma$ for $\bar B \rightarrow D^{(*)}$ for the remaining Dirac structures $\Gamma = 1, \gamma^5, \sigma^{\nu\rho} $, labeled as S, P, T respectively. For $\bar B \rightarrow D$ they read:
\begin{align}
    T_S^{\mu} & = (w+1) \zeta_1 \left( \frac{v'^\mu}{k \cdot v'} - \frac{v^\mu}{k \cdot v} \right) + \zeta_2 \, \left( v^\mu k \cdot v' - v'^\mu k \cdot v \right) \nonumber \\
   -i T_P^{\mu} & = \left((w-1)\zeta_3 - \zeta_2\right) \epsilon^{\mu\nu\rho\sigma} k^\nu v^\rho v'^\sigma \nonumber \\
   -i T_T^{\mu\nu\rho} & = \zeta_1 \left( \frac{v'^\mu}{k \cdot v'} - \frac{v^\mu}{k \cdot v} \right)  v'^{[\nu} v^{\rho]} \nonumber \\ 
   &+ \zeta_2 \left( v^\mu v'^{[\nu} k^{\rho]} +v'^\mu v^{[\nu}k^{\rho]} + g^{\mu[\nu} \left( v'^{\rho]} k \cdot v +v^{\rho]} k \cdot v' - (w+1) k^{\rho]}\right) \right)\nonumber \\
   &+ \zeta_3 \left[ v'^\mu \left((v' - w v)^{[\nu} k^{\rho]} - (k \cdot v) v'^{[\nu} v^{\rho]}\right) + v^\mu \left((v - w v')^{[\nu} k^{\rho]} - (k\cdot v')v^{[\nu} v'^{\rho]}\right) \right. \nonumber \\
   & + g^{\mu [\nu} \left( k \cdot v ( v- w v')^{\rho]} + k\cdot v' (v' - w v)^{\rho]} + (w^2-1) k^{\rho]}\right) \nonumber \\
   & + \zeta_4 \left( (v+v')^\mu k^{[\nu} (v - v')^{\rho]}  -(k \cdot ( v+ v')) g^{\mu[\nu}(v-v')^{\rho]} \right).
\end{align}
and for $ \bar B \rightarrow D^*$:
\begin{align}
    T_S^{*\,\mu} & = \zeta_2 \epsilon^{\mu\nu\rho\sigma} k_\nu (v+v')_\rho \varepsilon^*_{D^*\,\sigma} - \zeta_3 ( v \cdot \varepsilon^*_{D^*} ) \epsilon^{\mu\nu\rho\sigma}k_\nu v_\rho v'_\sigma \nonumber \\
    & + \zeta_4 k^{\nu}(v+v')^{[\nu} \epsilon^{\mu]\rho\sigma\delta}v_\rho v'_\sigma \varepsilon^*_{D^*\,\delta}  \nonumber \\
   -i T_P^{*\,\mu} & = \zeta_1 \left( \frac{v'^\mu}{k \cdot v'} - \frac{v^\mu}{k \cdot v} \right) v \cdot \varepsilon^*_{D^*} + \zeta_2 k^\nu  (v-v')^{[\mu} \, \varepsilon^{*\,\nu]}_{D^*} \nonumber \\
   & + \zeta_4 \, k^\nu \left( v^{[\mu} v'^{\nu]} v\cdot \varepsilon^*_{D^*} - (w-1) (v+v')^{[\mu} \varepsilon^{*\,\nu]}_{D^*}\right) \nonumber \\
   -i T_T^{*\,\mu\nu\rho} & = \zeta_1 \left( \frac{v'^\mu}{k \cdot v'} - \frac{v^\mu}{k \cdot v} \right) \epsilon^{\nu\rho\sigma\lambda}(v+v')_\sigma \varepsilon^*_{D^*\,\lambda}  + \zeta_3\, \epsilon^{\mu\alpha\beta\delta} k_\alpha v_\beta v'_\delta \,(v-v')^{[\nu}\varepsilon^{*\,\rho]}_{D^*} \nonumber \\
   & + \zeta_2 k^\sigma \left(\varepsilon^{*\,[\mu}_{D^*} \epsilon^{\sigma]\nu\rho\lambda}(v+v')_\lambda + (v'-v)^{[\nu}\epsilon^{\rho]\mu\sigma\lambda}\varepsilon^{*}_{D^*\,\lambda} \right) \nonumber \\
   & + \zeta_4  k^\sigma \varepsilon^*_{D^*\,\lambda} \left[(v+v')^{[\mu} \left( g^{\sigma][\nu}\epsilon^{\rho]\alpha\beta\lambda}v_{\alpha} v'_\beta  + \epsilon^{\sigma]\nu\rho\lambda}  -\epsilon^{\sigma]\alpha\lambda[\nu}v^{\rho]}v'_\alpha\right) + v'^{[\mu} v^{\sigma]} \epsilon^{\nu\rho\alpha\lambda}v'_\alpha\right].
\end{align}

\bibliographystyle{apsrev4-1}
\bibliography{bibliography}

\end{document}